\documentclass[amsmath,amssymb,11pt,superscriptaddress,reprint, preprintnumbers, notitlepage,aps,prl,twocolumn]{revtex4-1}

\pdfoutput=1 

\usepackage{graphicx}  
\usepackage{dcolumn}   
\usepackage{bm}        
\usepackage{amssymb,amsfonts,amsmath,physics}   
\usepackage{slashed}

\usepackage{tikz-feynman}
\tikzfeynmanset{compat=1.1.0}
\usetikzlibrary{arrows}
\tikzset{
    vertex/.style={circle,draw, minimum size=1.5em},
    edge/.style={->,> = latex'}
}

\usepackage[bookmarks, breaklinks, colorlinks,urlcolor=blue, citecolor=red, 
linkcolor=blue]{hyperref}

\usepackage[normalem]{ulem}

\def\to {\rightarrow}

\newcommand{\bmt}{\begin{pmatrix}}
\newcommand{\emt}{\end{pmatrix}}
\newcommand{\ba}{\begin{array}{c}}
\newcommand{\ea}{\end{array}}
\newcommand{\be}{\begin{equation}}
\newcommand{\ee}{\end{equation}}
\newcommand{\bea}{\begin{eqnarray}}
\newcommand{\eea}{\end{eqnarray}}

\newcommand{\bi}{\begin{itemize}}
\newcommand{\ei}{\end{itemize}}

\newcommand{\baz}{\begin{array}{cc}}
\newcommand{\mathsym}[1]{{}}

\newcommand{\bt}{\begin{tabular}}
\newcommand{\et}{\end{tabular}}

\newcommand{\benu}{\begin{enumerate}}
\newcommand{\eenu}{\end{enumerate}}

\def\mdm{m_{\rm DM}}

\def\lcal{\mathcal{L}}
\begin{document}
\title{Mono-X signal and two component dark matter: new distinction criteria}

\author{ Subhaditya Bhattacharya}
\email[Email:]{subhab@iitg.ac.in}
\affiliation{Department of Physics, Indian Institute of Technology Guwahati, Assam-781039, India}

\author{Purusottam Ghosh}
\email[Email:]{spspg2655@iacs.res.in}
  \affiliation{School Of Physical Sciences, Indian Association for the Cultivation of Science, 2A and 2B, Raja S.C. Mullick Road, Kolkata 700032, India} 

\author{ Jayita Lahiri}
\email[Email:]{jayita.lahiri@desy.de} 
   \affiliation{II. Institut f{\"u}r Theoretische Physik, Universit{\"a}t Hamburg, Luruper Chaussee 149, 22761 Hamburg, Germany}

\author{ Biswarup Mukhopadhyaya} 
\email[Email:]{biswarup@iiserkol.ac.in} 
  \affiliation{Department of Physical Sciences, Indian Institute of Science Education and Research Kolkata, Mohanpur - 741246, India}


\begin{abstract} 
\noindent
The identification and isolation of two WIMP dark matter (DM) components at colliders is of wide interest on the one hand but extremely 
challenging on the other, especially when the dominant signal of both DM components is of the mono-X type ($X=\gamma, Z, H$). After 
emphasizing that an $e^+e^-$ collider is more suitable for this goal, we first identify the theoretical principles that govern the occurrence 
of two peaks in missing energy ($\slashed{E}$) distribution, in a double-DM scenario. We then identify a variable that rather spectacularly elicits the 
double-peaking behaviour, namely, the plot of bin-wise statistical significance ($S/\sqrt{B}$) against $\slashed{E}$. Using Gaussian fits of 
the histograms, we apply a set of criteria developed by us, to illustrate the above points numerically for suitable benchmarks.
\end{abstract}

\maketitle



\noindent
Given that the visible sector of the universe comprises of multiple particle components at various mass scales, 
having more than one weakly interacting massive particles (WIMP) as DM components is quite viable~\cite{Cao:2007fy,Bhattacharya:2013hva,Herrero-Garcia:2018qnz,Aoki:2018gjf,Belanger:2021lwd,Hernandez-Sanchez:2022dnn}
and intriguing. Moreover, multicomponent DM frameworks provide some interesting dynamics and phenomenology like manifesting modified freeze-out, 
relaxing direct and indirect search bounds, addressing structure formation etc.

Multi-component DM's should in general lead to missing energy ($\slashed{E}$) or missing transverse energy ($\slashed{E_T}$) 
at colliders. However, distinguishing two DM components 
in collider experiments is rather challenging  in general, and is possible under certain circumstances only. The task is especially arduous if the major collider 
signal consists in mono-X plus $\slashed{E}$ (with X = $\gamma, Z, h$), primarily due to the very limited number 
of kinematic variables at our disposal, and partially due to the apparent insurmountability of Standard Model (SM) backgrounds. 
The present work is aimed at meeting such challenges.

$e^+ e^-$ colliders may be better suited for isolating two WIMP DM particles if they are kinematically accessible. 
This is mainly due to (a) the possibility of using $\slashed{E}$ as a variable, which depends on DM mass, 
(b) availability of missing longitudinal momentum as a source of extra information, 
(c) less SM background contamination, and (d) the advantage of initial beam polarisation~\cite{Fujii:2018mli,Habermehl:2020njb}. The rest of the discussion 
is therefore in the context of $e^+ e^-$ machines. We illustrate the criteria for double-DM signals being identifiable in $e^+e^-$ colliders, using some 
phenomenologically selected benchmarks. In particular, we suggest a variable that rather spectacularly
enables the distinction between the two DM components, namely, the bin-wise signal significance in
the missing-energy distribution.

{\bf Challenge 1: Will there always be noticeable peaks/bumps?} 
Mono-X signals are generic to any WIMP model, which result from direct production of a DM pair 
(when the DM is stabilised by a symmetry such as $Z_2$) typically in absence of an extended dark sector~\cite{Bell:2012rg,Berlin:2014cfa,Tolley:2016lbg,CMS:2017qyo,Bernreuther:2018nat,Bottaro:2021snn}
. 
It has the simplest event topology, and at the same time leaves 
one with limited choice in analysing the final state kinematics. Bumps in $\slashed{E}$ distributions 
are driven by both kinematics and dynamics. Kinematics decide the 
minimum and maximum $\slashed{E}$ for a given event, which in turn depend on the centre-of-mass (CM) energy and 
DM masses. Energy-momentum conservation plays an important role here. On the other hand, the shape 
of the $\slashed{E}$ distribution, both at the edges and in between, depends on the dynamics, namely, the Lagrangian, 
including possible sources of soft or collinear divergence. 

In order to explore the features of two-component DM in a model-independent manner we take up an Effective Field Theory 
(EFT) approach as it suffices to address the DM saturation as well as detectability.
Table~\ref{table_operators} shows a set of EFT operators 
involving a iso-singlet DM ($\chi$), stabilised by a $Z_2$ symmetry, interacting with SM particles which can lead to 
$e^+ e^- \rightarrow\chi\chi X$ ($X= \gamma, Z, H$)~\cite{Fox:2011fx,Chae:2012bq,Kundu:2021cmo,Barman:2021hhg}. 
This can happen via either $2 \rightarrow 3$ hard scattering, or 
through $2 \rightarrow 2$ DM pair production along with radiation of a $\gamma$, $Z$ or $H$ from one of the legs. 
The benchmarks on which we base our illustrative numerical predictions include dual-DM scenarios 
with both scalars, fermions and combinations of them. Note further that the operators listed here are merely 
indicative, and by no means constitute an exhaustive list. However, whether the bump-hunting approach works 
for a particular operator depends on the following principles:

\begin{table}[!hptb]
\begin{center}
\begin{tabular}{| c | c | c |}
\hline
 DM-type & Name &  Operator  \\
\hline
\hline
Scalar & $O_1^{s}$ & $\frac{c}{\Lambda^2} (\bar{L} \Phi \ell_R) (\chi^2)$ \\
\hline
Scalar & $O_2^{s}$ & $\frac{c}{\Lambda^4} (\bar{L} D_{\mu} \Phi \ell_R) (\chi \partial_{\mu} \chi)$ \\
\hline
Scalar & $O_3^{s}$ & $\frac{c}{\Lambda^2} (B_{\mu\nu}B^{\mu\nu} + W_{\mu\nu}W^{\mu\nu})(\chi^2)$ \\
\hline 
\hline
Fermion & $O_1^{f}$& $\frac{c}{\Lambda^2} (\bar L \gamma^{\mu} L  + \bar \ell_R \gamma^{\mu} \ell_R \bar)(\bar \chi\gamma_{\mu} \chi)$ \\
\hline
Fermion & $O_2^{f}$ & $\frac{c}{\Lambda^3} (\bar L \Phi \ell_R) (\bar \chi \chi)$ \\
\hline
Fermion & $O_3^{f}$ & $\frac{c}{\Lambda^3} (B_{\mu\nu}B^{\mu\nu} + W_{\mu\nu}W^{\mu\nu})(\bar \chi \chi)$ \\
\hline
\end{tabular}
\caption{Samples of DM-SM EFT operators involving isosinglet DM ($\chi$) as scalar or fermion. $B_{\mu\nu}$ and $W_{\mu\nu}$ are electroweak field strength tensors, 
$L$($\ell_R$) are left (right)-handed SM lepton doublet (singlet), $\Phi$ is the SM Higgs doublet, 
$D_{\mu}$ is the gauge-covariant derivative, $c$ is the Wilson coefficient and $\Lambda$ is the cut-off scale of the EFT operators.}
\label{table_operators}
\end{center}
\end{table}

\begin{enumerate}
\item Angular momentum conservation: This is often related to to the chiral structure of the interactions involved.
There can be situations where a particular spatial configuration and the relative directionality 
of the particles is not allowed by the angular momentum conservation, given the helicity assignments demanded by the underlying 
interactions. In that case we see a drop in the differential distribution in $\frac{d\sigma}{d{\slashed{E}}}$ or zeroes at the edge(s) of the phase space, 
providing a peaking behaviour.
\item Rotational invariance in Minkowski space ($E^2 - |p|^2 = m^2$): The last integrand before arriving at the differential distribution 
such as $\frac{d\sigma}{d{\slashed{E}}}$, always contains a positive power of the of three-momentum of a final state particle. 
Therefore the final state particles being at rest automatically causes the differential distribution to vanish resulting a peak-like behaviour, 
unless the drop is assuaged by a factor in the denominator, where the same three-momentum occurs. 
\item Collinear divergence: Arises when the radiated $\gamma$ is collinear to the initial or final state particles and dominantly controls 
features of the distributions. A massive propagator which does not lead to collinear divergence, may still lead, instead of a drop, to a rise 
at an edge of the phase space due to a three momentum factor in the denominator. A soft photon corresponding to the maximum of $\slashed{E}$, 
accompanied by an infrared divergence, may also cause such a rise.
\item Operator structure: The exact position of the maxima of the distributions are often dictated by the specific interaction in question.  
\item DM mass ($\mdm$): The left-edge of the $\slashed{E}$ distribution, i.e. the minimum possible $\slashed{E}$ 
depends on $\mdm$. Sizes of the peaks depend on the production cross-section, which in 
turn also depends on $\mdm$ together with $c$ and $\Lambda$.
\end{enumerate}  

\begin{figure}[!hptb]
	\includegraphics[width=6.5cm,height=4.2cm]{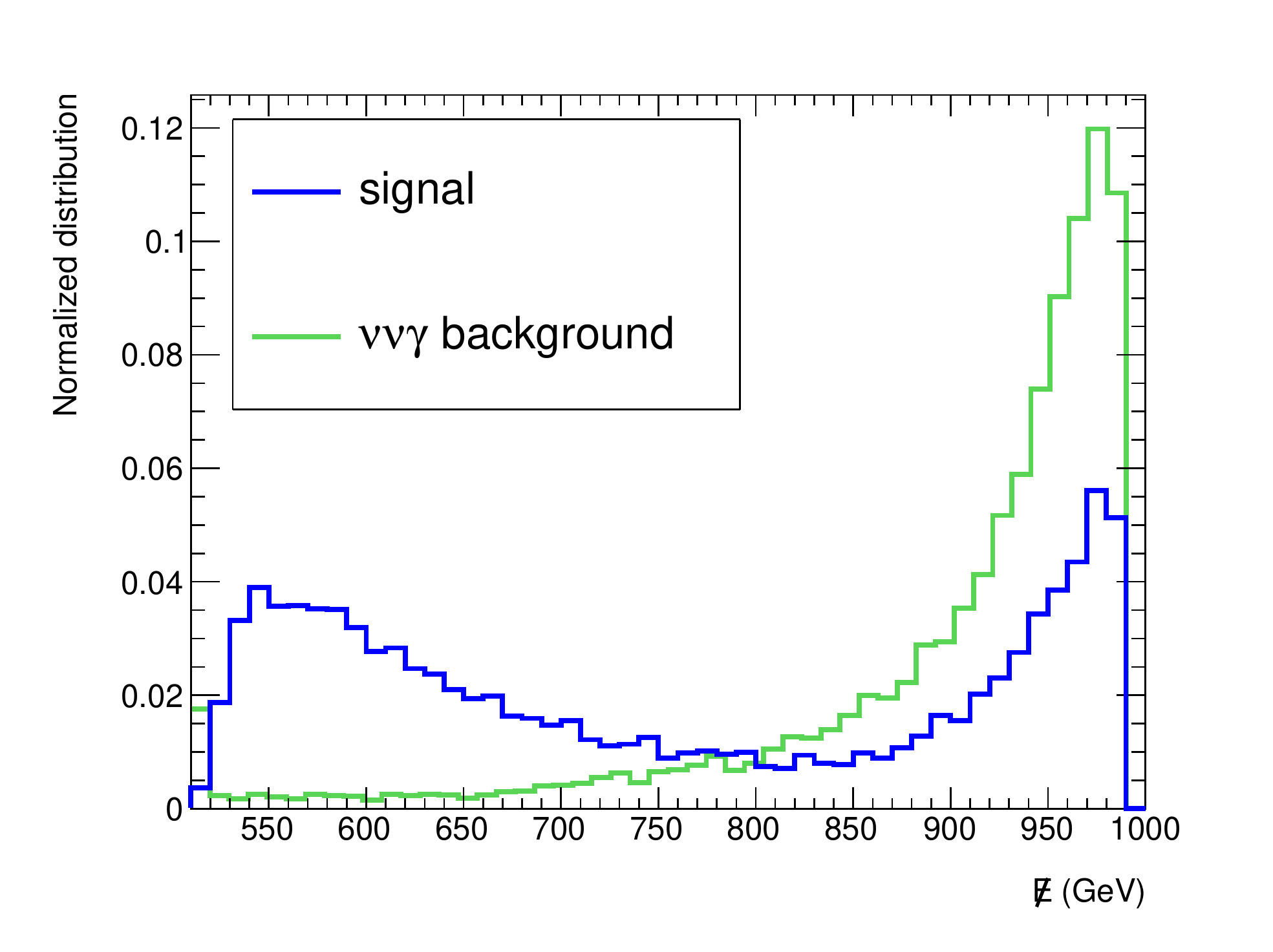} 
	\includegraphics[width=6.5cm,height=4.2cm]{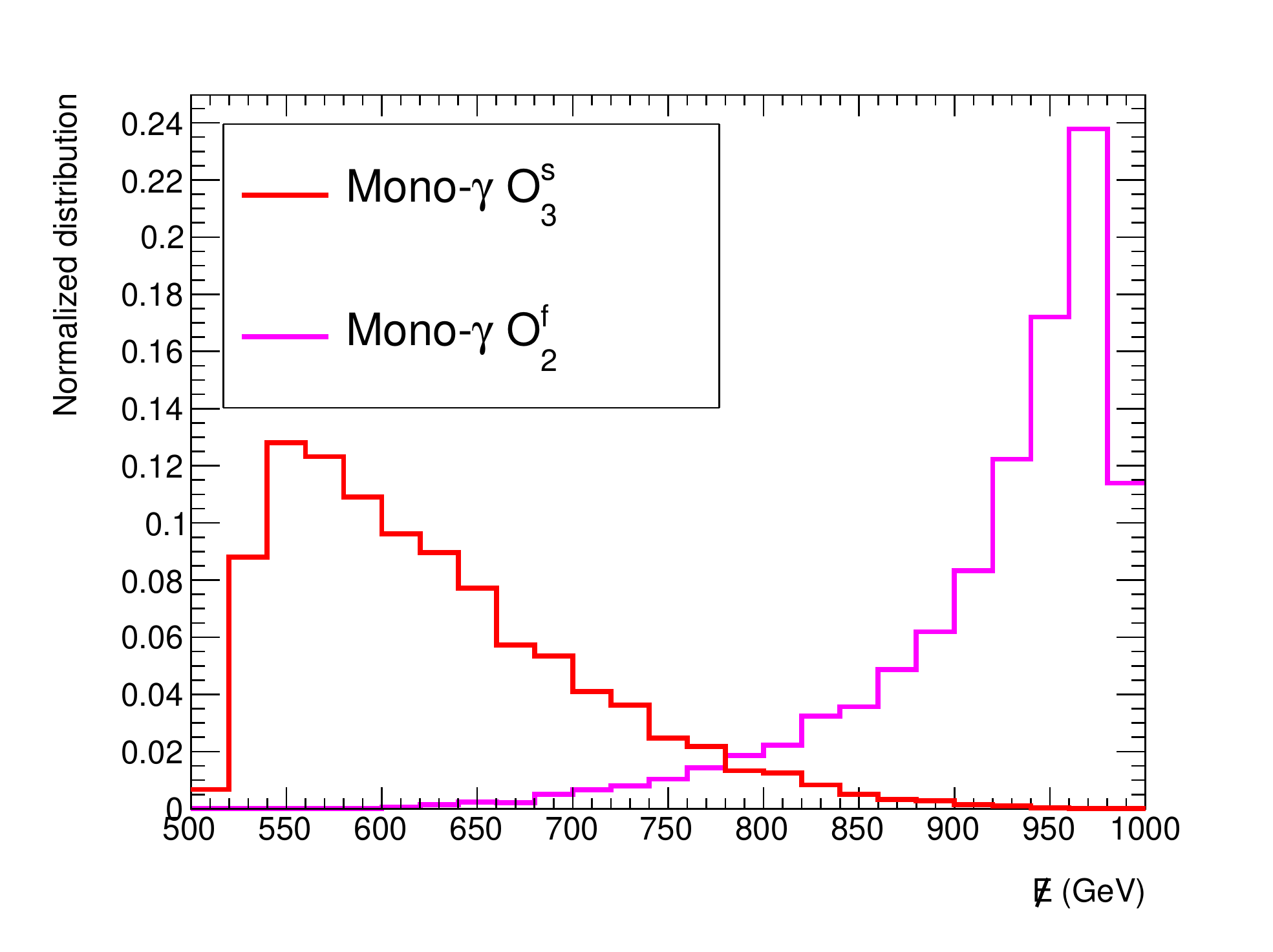}
	\caption{[Top] Normalised $\slashed{E}$ distribution in the mono-$\gamma$ final state for two DM components arising from $O_3^s$ and $O_2^f$ 
	operators, for DM masses 100 GeV and 200 GeV respectively, together with $\nu\bar\nu \gamma$ background. 
    [Bottom] Individual normalized distributions of $\slashed{E}$ in the mono-$\gamma$ final state coming from the two DM candidates 
	of masses 100 GeV ($O_3^{s}$) and 200 GeV ($O_2^{f}$). For both figures, we choose $\sqrt{s} = 1$ TeV, $\Lambda=1.2$ TeV with $P_{e^{+}} = +30\%, P_{e^{-}} = +80\%$.}
	\label{two_comp_monophoton}
\end{figure}

\begin{figure}[!hptb]
     \includegraphics[width=6.5cm,height=4.2cm]{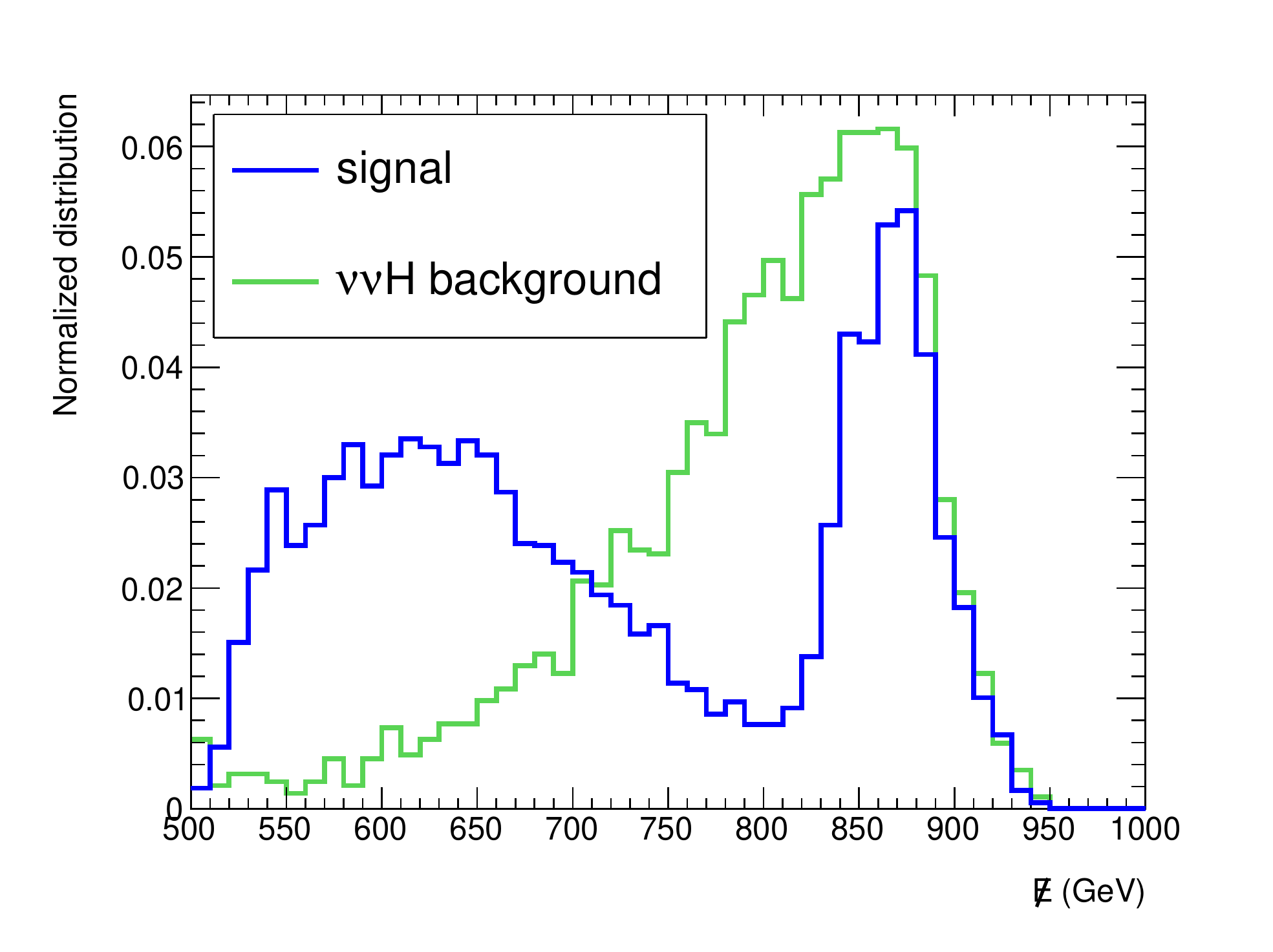} 
	\includegraphics[width=6.5cm,height=4.2cm]{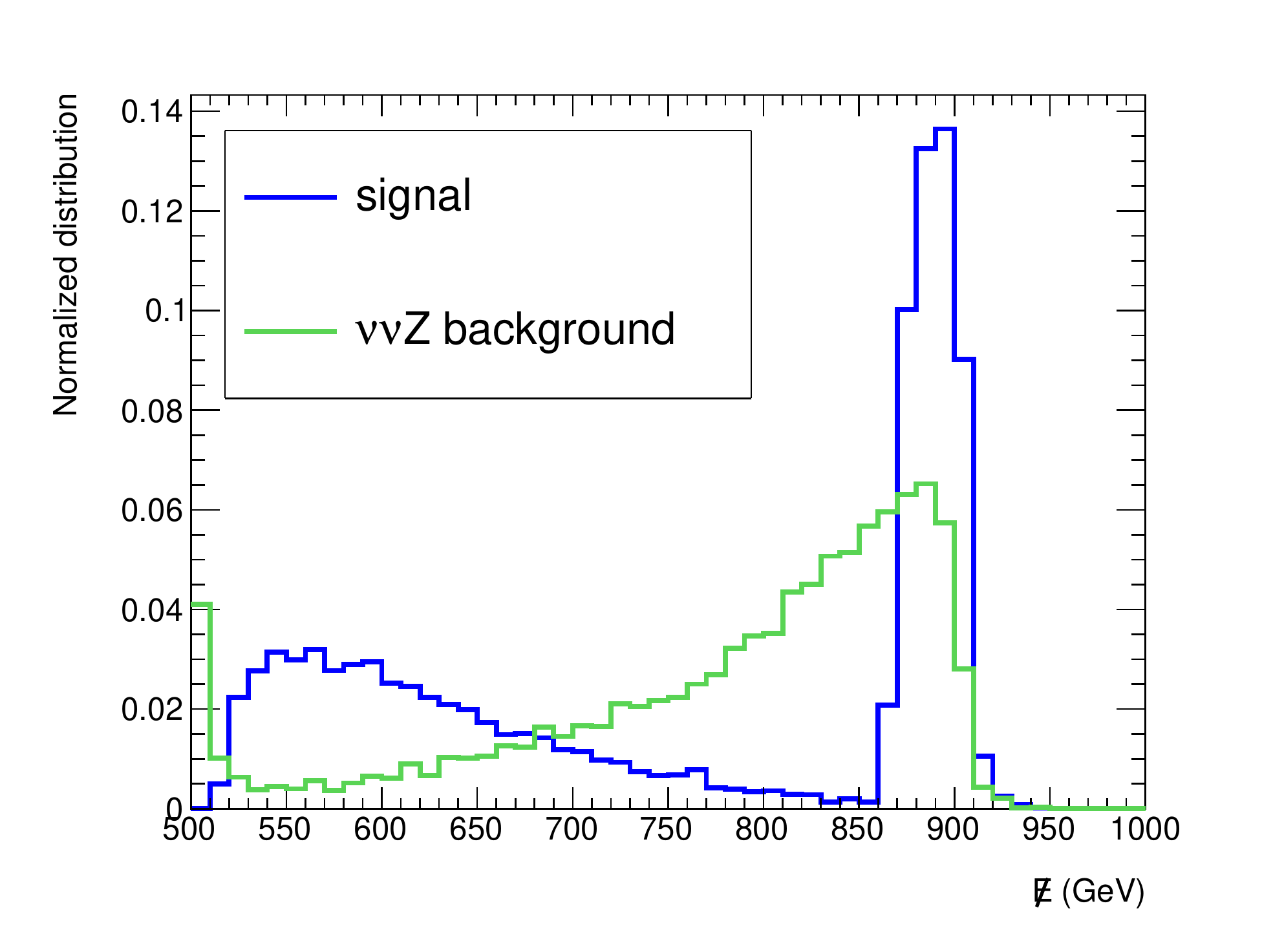}
	\caption{Normalised $\slashed{E}$ distribution for (top)  mono-$H$ final state for two DM components from $O_2^s$ and $O_1^s$, (bottom) mono-$Z$ final state for two DM components arising from $O_3^s$ and $O_1^f$, with masses 100 GeV and 400 GeV respectively. 
	SM backgrounds are also shown. We choose $\sqrt{s} = 1$ TeV, $\Lambda=1.2$ TeV and 
	$P_{e^{+}} = +30\%, P_{e^{-}} = +80\%$.}
	\label{two_comp_monoz_monoh}
\end{figure}

Our study is centred around double bump hunting in $\slashed{E}$ distributions coming from two DM components. 
It is worth mentioning that, when two DM-components stem from the same operator having same spin, 
the higher mass DM having smaller production cross-section gets hidden beneath the tail of the larger peak, 
failing to produce a double-peak behaviour, see the supplemental material for example.

In Figure~\ref{two_comp_monophoton}(top) we show an example of $\slashed{E}$ distributions in mono-$\gamma$ signal, 
where double peak behaviour is observed, when two DM particles from two different operators $O_3^s$ and $O_2^f$
are pair-produced in $e^+ e^-$ collision at $\sqrt{s} = 1$ TeV. The peaking behaviour around the minimum of 
$\slashed{E}$ is obtained because ($i$) there is no collinear divergence at this kinematic point, and ($ii$) 
this region is not disfavoured by angular momentum conservation, thanks to the chiral structure 
of the interaction responsible for this peak ($O_3^s$). $O_3^s$ provides a photon mediated final state consisting of a photon and two scalar DM's.
The resultant spin one state provides the minimum $\slashed{E}$ configuration when the photon emerges opposite to the two DM particles.
The rise at the maximum of $\slashed{E}$, on the other hand, is because 
of collinear as well as infrared divergence, since the $\gamma$ emitted is not only `soft' but also at a 
small angle with the initial $e^{\pm}$, something that in this case is not disfavoured by the angular momentum conservation. 
This is clear from Figure~\ref{two_comp_monophoton}(bottom), where we show the individual contributions from the two DM components 
($O_3^s$ in red and $O_2^f$ in purple) to the $\slashed{E}$ signal distribution as shown by the blue line in top.

\begin{figure}[!hptb]	
\includegraphics[scale=0.21]{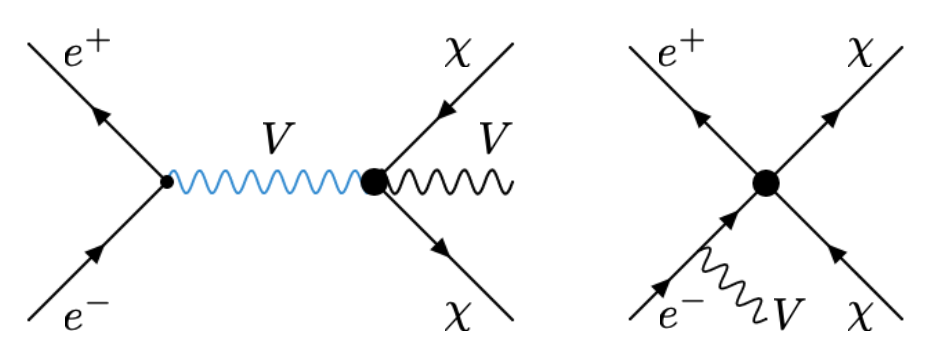}
\caption{Feynman graphs contributing to mono-$V$ ($V=\gamma,H,Z$) events in this analysis.}
\label{fig:fd}
\end{figure}

We note here that the double peaking behaviour can also arise from a single DM
component, when $\mdm$ is very small compared to $\sqrt{s}$, since
a collinear divergence can then occur even at the minimum of $\slashed{E}$ or when 
it interacts via different operators, as shown in the supplemental material. On the other hand, 
the combined $\frac{d\sigma}{d{\slashed{E}}}$ distributions for two DM particles from two different operators  
can provide a single peak, when the individual distributions are identical and shows a collinear shoot-up at the maximum 
$\slashed{E}$ and a suppression at the minimum $\slashed{E}$ due to the structure of relevant operators. 
This is avoidable if one of the two mono-$\gamma$ generating processes is driven by an effective 
interaction vertex as chosen above, as opposed to the case where the photon is emitted via bremsstrahlung. 
 
 Next, we take up the mono-$H$ and mono-$Z$ signals, illustrative predictions for which are shown in Figure \ref{two_comp_monoz_monoh}. 
$O_2^s$ and $O_1^s$ for mono-$H$ (top panel), $O_3^s$ and $O_1^f$ are considered for mono-$Z$ (bottom panel)
with details of the parameters stated in the caption. We have reconstructed the mono-$Z$ and 
mono-$H$ events in the $\ell^+ \ell^- (\ell = e, \mu)$ and $b\bar{b}$ channel, respectively, with the 
standard trigger and selection cuts. For both the plots in Figure \ref{two_comp_monoz_monoh}, 
the distribution is brought down to zero for maximum $\slashed{E}$. The reason behind this being the 
occurrence of $(E^2 - m^2)$ in the numerator of the penultimate integrand in cross-section calculation. 
This factor, traceable to rotational invariance in the Minkowski space, causes the distribution to vanish at the maximum of $\slashed{E}$, which corresponds
to zero three-momentum of the $Z$ or the $H$. This is conducive to peaking behaviour corresponding to each the
two DM components, so long as the behaviour of low $\slashed{E}$ is favoured by angular momentum
conservation, especially since there is now no collinear divergence at the higher end. One 
may thus see two peaks, the level of shift to the left for each peak being governed by the 
corresponding DM mass; their splitting thus becomes the deciding factor in the isolation of the peaks. 
An exceptional case is the operator $O_1^f$ where the $Z$ can be emitted via bremsstrahlung, 
for which the peak occurs at the higher end of $\slashed{E}$ spectrum for any DM mass due to 
the appearance of a small number $\sim m^2_Z/s$ in the denominator of the penultimate integrand.
Feynman graphs that contribute to mono-$V$ events ($V=\gamma, H, Z$) considered in this analysis are 
shown in Figure. \ref{fig:fd}, that are typical to result double bump in $\slashed{E}$ distributions.

\begin{figure}[!hptb]
\includegraphics[scale=0.31]{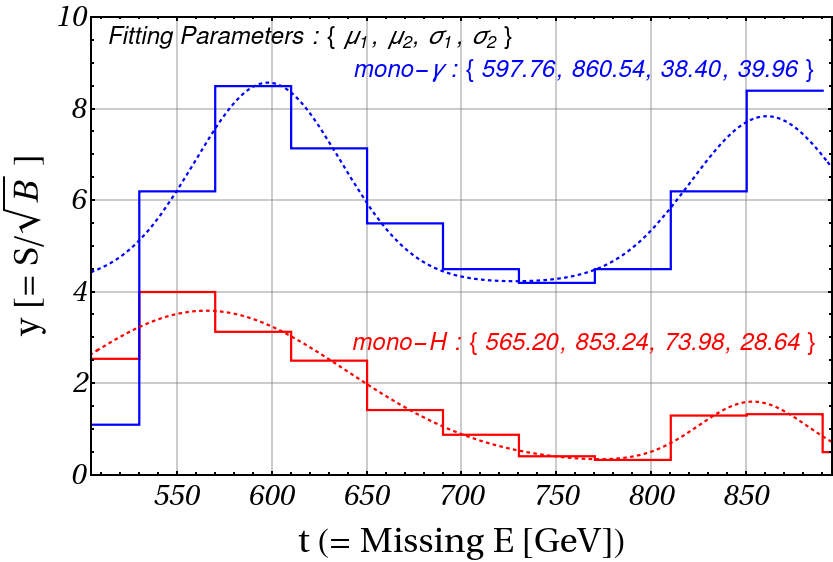} \\ 
\caption{ Variation of local signal significance and Gaussian fitting as a function of $\slashed{E}$ for mono-$\gamma$ signal (blue lines) with $\lcal=$100 $\rm fb^{-1}$ as in Figure \ref{two_comp_monophoton} (top). The same are plotted for mono-$H$ signal (red lines) as in Figure~\ref{two_comp_monoz_monoh} (top) at $\lcal=$1000 $\rm fb^{-1}$. Global signal significances = 5$\sigma$ and $S/B \approx$ 4\% for mono-$\gamma$. For mono-$H$:  4.2$\sigma$ and $S/B \approx$ 4\%. Interestingly, in this case the signal-only distribution does not show a discernible double-peak behaviour, but the bin-wise significance does, as shown here.}
\label{o4f_o2f_sigbkg_monophoton}
\end{figure}
\begin{figure}[!hptb]	 
 \includegraphics[scale=0.31]{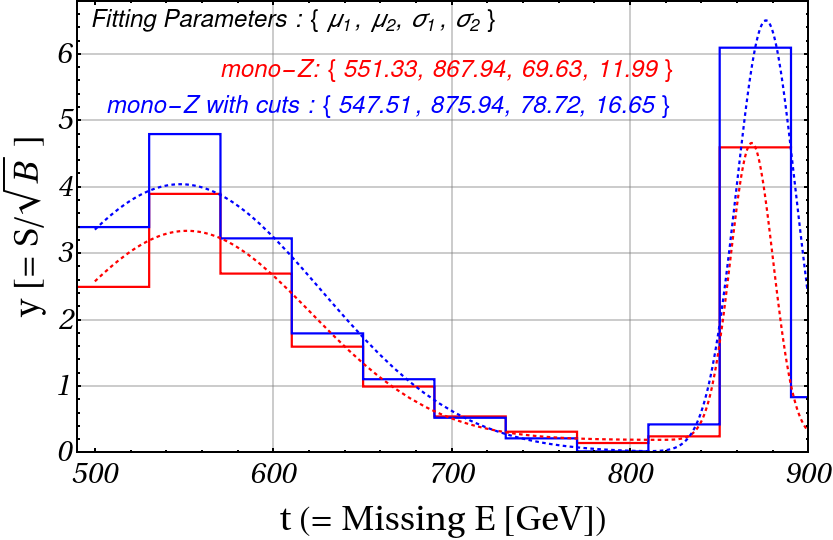} 
 \caption{ Variation of local signal significance and Gaussian fitting as a function of $\slashed{E}$ for mono-$Z$ signal (red lines) as 
 in Figure~\ref{two_comp_monoz_monoh} (bottom) at $\lcal=$1000 $\rm fb^{-1}$. The same are plotted for the mono-$Z$ signal after applying the cut $\slashed{E_T} < 50$ GeV or $> 200$ GeV at $\lcal=$ 1000 $\rm fb^{-1}$ (blue lines). Global signal significance $4.7 \sigma$ and $S/B \approx$ $5\%$ (red line). After applying the cuts, $S/B$ increased to approximately $16\%$, and the global significance reached $7\sigma$.}
\label{o1f_o3s_sigbkg_monoz}
\end{figure}

We have used MADGRAPH\cite{Alwall:2011uj} in tandem with DELPHES with the ILD card~\cite{deFavereau:2013fsa},
as a result of which the behaviour seen in the distributions deviate a little bit from the
usual ones. Specific beam polarisation ($P_{e^{+}} = +30\%, P_{e^{-}} = +80\%$) choices are made to optimise the relative 
heights of both the peaks.  

We also note that with appropriate combination of $c$, $\Lambda$ and $\mdm$, it is possible to achieve the observed relic density from 
two DM-components constituted of the EFT operators in Table \ref{table_operators}, 
while addressing collider distinguishability, see the supplemental material. \\

\noindent
{\bf Challenge 2: How to make the double peaks rise above the backgrounds?} We have seen 
above some examples of cases where two peaks/bumps corresponding to the two DM particles are
noticeable. However, it is not obvious that the peaks are distinct when different kinds of $\nu {\bar{\nu}} X 
(X = \gamma, Z, H)$ backgrounds as already shown in Figures \ref{two_comp_monophoton} and \ref{two_comp_monoz_monoh}, 
are taken into account. At $e^+e^-$ collider with $\sqrt{s} = 1$ TeV and polarizations $P_{e^{+}} = +30\%$ and $P_{e^{-}} = +80\%$, the corresponding SM backgrounds are as follows: $674.0$ fb for $\nu\nu\gamma$, $46.6$ fb for $\nu\nu H(b\bar b)$, and $18.0$ fb for $\nu\nu Z(\ell^+\ell^-)$.

While we consider all possible backgrounds, the most menacing ones that tend to drown the mono-$Z$ and 
mono-$H$ signals is the one from vector boson fusion (VBF). There are, after all, not too many kinematic variables
at our disposal in this case. How then can one rise above the irreducible backgrounds? 

Two generic criteria for this to happen, securing the best signal-to background ratio, are: 
(i) the signal peak that is closest to the SM background peak should rise as much above the latter as possible, 
and (ii) the other signal peak, after any distortion by the background, should still be as distinct from the first one as is possible.

In order to make the double peak distinct in spite of backgrounds, we propose a new observable here. This is the bin-by-bin signal significance 
$S/\sqrt{B}$ for an integrated luminosity ($\lcal$) optimised for different DM benchmarks. 
This `local significance', plotted against $\slashed{E}$, is shown 
in Figures \ref{o4f_o2f_sigbkg_monophoton} (for mono-$\gamma$ and mono-$H$), \ref{o1f_o3s_sigbkg_monoz} (for mono-$Z$). We present alongside 
the Gaussian fits of the corresponding histograms after minimizing $\chi^2$, using:
 \bea\label{eq:gf}
 G(t)&=& G_1(t)+G_2(t) + \cal{B} \nonumber\\
 &=& A_1~ e^{-\frac{(t-\mu_1)^2}{2\sigma_1^2}}+A_2 ~ e^{-\frac{(t-\mu_2)^2}{2\sigma_2^2}} + \cal{B} ~.
 \eea
The values of $\mu_1$, $\mu_2$, $\sigma_1$ and $\sigma_2$ in each case are indicated in the figure insets. 
In the examples shown here, we have taken the bin size along the $\slashed{E}$-axis to be 40 GeV, which in most of cases 
optimise the significance. The central outcome of this is a rather remarkable distinctiveness of the two DM peaks, 
as evinced from the two Figures \ref{o4f_o2f_sigbkg_monophoton}, \ref{o1f_o3s_sigbkg_monoz}. The global significance and the overall $S/B$ are given in the
corresponding figure captions; one can notice $S/B$ ratios ranging upto 10\%,
and the global significance ranging between $4-5\sigma$. This not only assures one
of the usefulness of the suggested distributions but also indicates the level of refinement
required in an $e^+ e^-$ experiment in connection with the reduction of systematic errors.

\begin{table*}{\footnotesize
\begin{ruledtabular}
\begin{tabular}{|c|c|c|c|c|c|c|c|c|}
 {\bf Benchmarks in} &{Operators}& $\{ m_{\rm DM1},~m_{\rm DM2} \} ({\rm GeV})$ &  $\{ \sigma_{\rm DM1},~\sigma_{\rm DM2} \} ({\rm fb})$ & $\mathcal{L} ({\rm fb^{-1}})$ &$S/B$ & $S/\sqrt{B}$(Global)& $R_{C3}$ & $R_{C4}$    \\
\hline \hline
Figure~\ref{o4f_o2f_sigbkg_monophoton}({mono-$\gamma$}) & $O_3^f, O_2^f$  & $\{100,~200\}$ & \{2.9,9.9\} & 100 & $4\%$ & $5.0~\sigma$ &  4& 1.8  \\
\hline
Figure~\ref{o4f_o2f_sigbkg_monophoton}({mono-$H$}) & $O_2^s, O_1^s$  & $\{100,~400\}$ & \{0.4,0.23\} & 3000 & $4\%$ & $4.2~\sigma$ & 38 & 2.1  \\
\hline
Figure~\ref{o1f_o3s_sigbkg_monoz}({ mono-$Z$}) & $O_3^s, O_1^f$  & $\{100,~400\}$ & \{0.37,0.42\} & 1000 & $5\%$ & $4.7~\sigma$ & 16 & 7.1  \\
\hline
Figure~\ref{o1f_o3s_sigbkg_monoz}({ with cuts}) & $O_3^s, O_1^f$  & $\{100,~400\}$ & \{0.37,0.42\} & 1000 & $16\%$ & $7.0~\sigma$ & 23 & 23.8  \\
\end{tabular}
\caption{The combination of effective operators, signal significance and corresponding values of $R_{C3}$ and $R_{C4}$ for the chosen benchmarks, 
as discussed in Figures~\ref{o4f_o2f_sigbkg_monophoton}-\ref{o1f_o3s_sigbkg_monoz}. $\sigma_{\rm DM(1,2)}$ denote the individual contribution to the total cross-section.}
\label{table_RC}
\end{ruledtabular}}
\end{table*}

Is there any way of further improving the discernibility of the two peaks? In response to this query,
we suggest further application of an $\slashed{E_T}$-cut on the signal and the backgrounds.
Since $\slashed{E_T}$ and $\slashed{E}$ are {\em correlated only partially}, and the $\mdm$
enters only into $\slashed{E}$ manifestly, one can, with judicious $\slashed{E_T}$ cuts, make the
two peaks more distinct, and at the same time improve the global significance and
the signal-to-background ratio. This is illustrated in Figure~\ref{o1f_o3s_sigbkg_monoz} by the blue solid line. 
It should be remembered, though, that the $\slashed{E_T}$-cuts need to be customised after noticing the double-peak
appearance first, lest one could either slice through the middle on a single peak and
lead to fake double-peak events or reduce the height of one of the peaks in an actual
double-peak event so that it effectively looks like a single peak.

The next question is: how to quantify the distinguishability of the two peaks in a mono-X signal?
For this purpose we refer the reader to reference~\cite{Bhattacharya:2022wtr} where, in the context of the double-peak events
generated in cascades, a number of distinction criteria $C_i (i = 1-4)$ have been developed; 
leading to the quantities $R_{C_i} (i = 1-4)$. The variables $R_{C_3}$, $R_{C_4}$ emerge as the most useful ones,
defined as, 
\bea
 R_{C3} = \frac{\int_{t_1^-}^{t_1^+} y dt-\int_{t_2^-}^{t_2^+} y dt}{\int_{t_1^-}^{t_1^+} y dt+\int_{t_2^-}^{t_2^+} y dt}\,,~
 R_{C4} = \frac{y(t^\prime) -y(t_{\rm min})}{\sqrt{y(t_{\rm min})}} \,;
\label{c3}
 \eea
 where $y=G(t)$, $t_i^\pm=t_i \pm \Delta t$ with $t_i$ denoting the mean value of $i^{\rm th}$ peak of the two peak gaussian function $G(t)$ and 
 $t_{\rm min}$ denotes the local minima between the two peaks. In the definition of $R_{C_4}$, the parameter $t^\prime$ is specified as 
 \bea\nonumber
 t^\prime=\left\{ 
  \begin{array}{ c l }
    t_2 & \quad \textrm{if } ~y(t_2) < y(t_1) , \\
    t_1 & \quad \textrm{if } ~y(t_2) > y(t_1) .
  \end{array}
\right.
\eea

$R_{C3}$ reflects a comparative estimate of the number of events in the vicinity of the two peaks, and smaller $R_{C3}$ implies 
that the second peak is more significant relative to the first one. $R_{C4}$ quantifies the presence of the peaks with 
respect to the minimum in between, and its bigger values indicate better distinguishability.
In Table \ref{table_RC}, we present the values corresponding to our chosen benchmark points, based on the Gaussian fits of the
significance histograms. We note further that for the chosen benchmarks, (a) Mono-$\gamma$ requires the least $\lcal$,
(b) Mono-$Z$ leads to the best distinguishability in terms of peak separation, and (c) Mono-$H$  
requires the maximum $\lcal$. The improvement brought about by the $\slashed{E_T}$
is also visible in Figure~\ref{o1f_o3s_sigbkg_monoz}.
Similarly three peaks in the $\slashed{E}$ distribution from three DM's can be addressed by fitting a three-peak Gaussian 
distributions and subsequent analysis.

\begin{figure}[!tbh]
    \centering
       \includegraphics[scale=0.35]{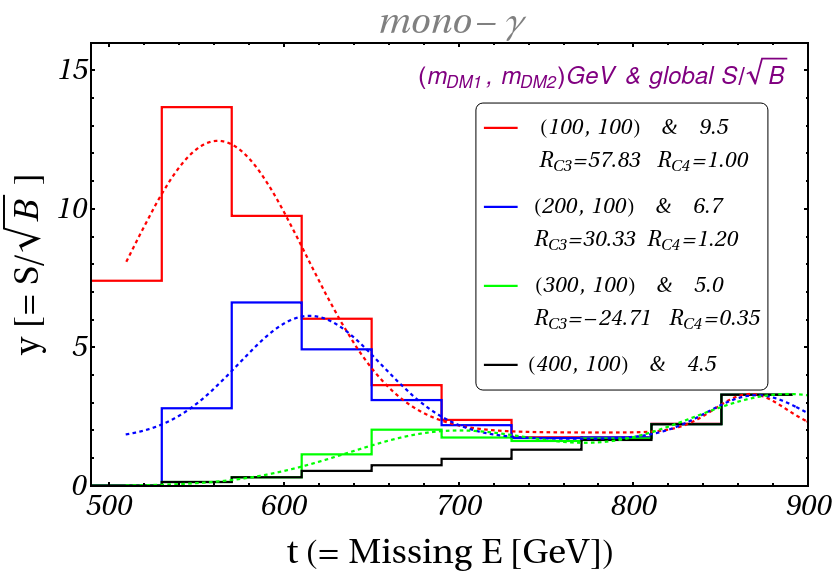}
    \caption{Variation of local signal significance and Gaussian fitting as a function of $\slashed{E}$ for mono-$\gamma$ signal 
for the operator set $O_3^s$ (DM1) and $O_2^f$ (DM2) with various combinations of $\mdm$ at $\lcal=$3000 $\rm fb^{-1}$. $R_{C_3}$, $R_{C_4}$ and global 
significance for each combination is mentioned in the inset.}
    \label{fig:diffMass2}
\end{figure}

 In Figure~\ref{fig:diffMass2}, we explore the effect of $\mdm$ in the distinguishability of the two peaks. We show the local $S/\sqrt{B}$-$\slashed{E}$ distribution 
for $O_3^s$ (DM1) and $O_2^f$ (DM2) for various choices of $\mdm$. It is clear that the distinguishability decreases with 
increasing $\rm m_{DM1}$ (see parameters $R_{C_3}$, $R_{C_4}$ in the inset), resulting a decline in the global significance (also mentioned in the inset) 
due to smaller production cross-section. For some other examples, see the supplemental material.

In conclusion, mono-X events in an $e^+ e^-$ collider bring in a big challenge in distinguishing between two DM components. 
This is in contrast to DM signals produced in cascades, where not only are backgrounds are easier to reduce, but the occurrence
of two peaks is also a more widespread phenomenon, as has been demonstrated in~\cite{Bhattacharya:2022wtr}. Our analysis of mono-X events identifies 
the physics underlying double-peak occurrence; we also demonstrate that bin-by-bin distributions
in signal significance plotted against $\slashed{E}$ obviates the double-peaking behaviour
which the backgrounds otherwise tend to obliterate. It is also found by us that  judicious 
$\slashed{E}_T$ cuts serve to enhance the prominence of the two peaks as well as overall significance
and signal-to-background ratios. We shall discuss in a more detailed paper other issues such as the 
demands on our sample theoretical scenario brought about by issues such as relic density or direct searches~\cite{multicoll1}.

One point is worth emphasizing at the end. There may be occasions where the DM signal actually yields a single peak due to one DM component,
but a fortuitous background pattern creates the semblance of a second peak. The bin-wise significance comes to the rescue, where one ends up with two-peak 
only at the presence of the two DM components provided they give rise to adequate signal rates individually and are made 
distinct through the variables discussed here.

~\\

{\bf Acknowledgments}: SB acknowledges the grant CRG/2019/004078 from SERB, Govt. of India. PG would like to acknowledge 
Indian Association for the Cultivation of Science, Kolkata for the financial support.

~\\~\\
\bibliographystyle{utphys}
\bibliography{Refs}
\begin{center}
 {\bf \large SUPPLEMENTAL MATERIAL}
\end{center}
\section{Two same type of DMs having different masses}
\label{sec1}
\begin{figure}[!hptb]
	\centering
	\includegraphics[width=6.5cm,height=5cm]{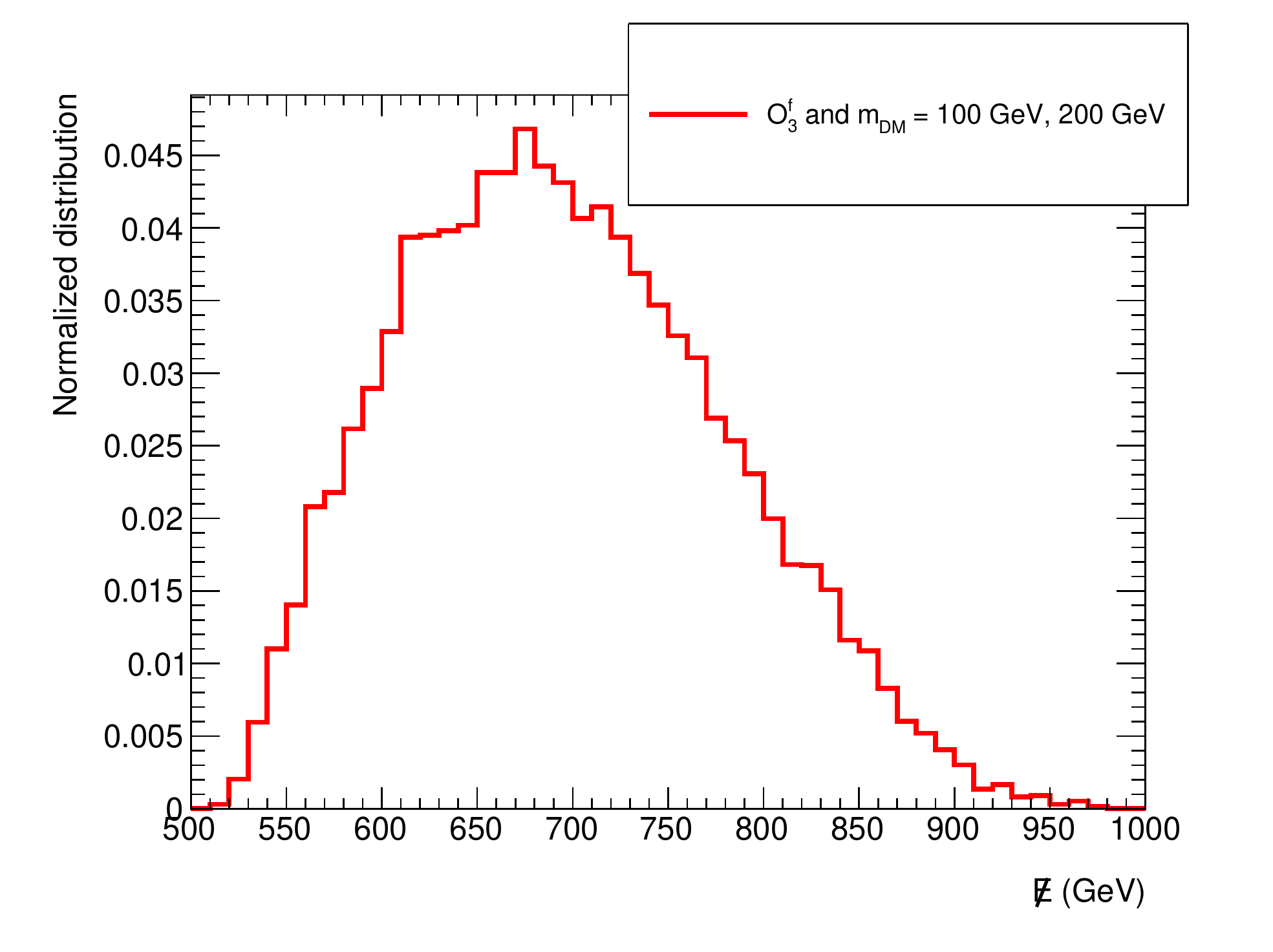}
       	\caption{Normalised $\slashed{E}$ distribution for the mono-$\gamma$ signal in $e^+e^-$ collider having two DM components ($\chi$) of 
	mass 100 GeV and 200 GeV, arising from the operator $O_3^f=\frac{1}{\Lambda^3} (B_{\mu\nu}B^{\mu\nu} + W_{\mu\nu}W^{\mu\nu})(\bar \chi \chi)$.
	We choose $\sqrt{s} = 1$ TeV, $\Lambda=1.2$ TeV with $P_{e^{+}} = +30\%, P_{e^{-}} = +80\%$. }
	\label{o3f_100_200}
\end{figure}

It is quite legitimate to have two distinct DM candidates of same type, i.e. with same spin but different masses, 
stemming from the same operator, stabilised by two different symmetries. They in principle can have separated $\slashed{E}$ distribution 
due to different masses. But in that case the higher mass DM will have much smaller production cross-section compared to the lighter DM. 
Since they both come from the same operator and is of same spin, there will be no other effect to compensate the reduction in cross-section. 
Therefore in that case, the peak with the smaller cross-section will always be hidden beneath the tail of the other larger peak, failing to produce 
a double-peak behaviour. To validate this point, we present here an example case where two fermion DM's ($\chi$) of mass 100 GeV and 200 GeV 
arise from a single operator $O_3^f=\frac{1}{\Lambda^3} (B_{\mu\nu}B^{\mu\nu} + W_{\mu\nu}W^{\mu\nu})(\bar \chi \chi)$ in Figure~\ref{o3f_100_200}. 
We see that the peak coming from 200 GeV DM is not seen at all, but gets submerged under the 100 GeV one. If the mass separation is small, then the 
cross-sections are similar, but the peaks are close by, so the indistinguishability remains. We must also remind here that this feature of indistinguishable peaks
for same type of DM arise for mono-X signal only, for extended dark sectors having produced heavy particles decaying to DM, the mass splitting rather play a 
more important role rather than the DM mass, so the double peak behaviour can be seen for same type of DM as well.

\section{Single DM interacting via different operators}

\begin{figure}[!hptb]
	\centering
	\includegraphics[width=7.5cm,height=6cm]{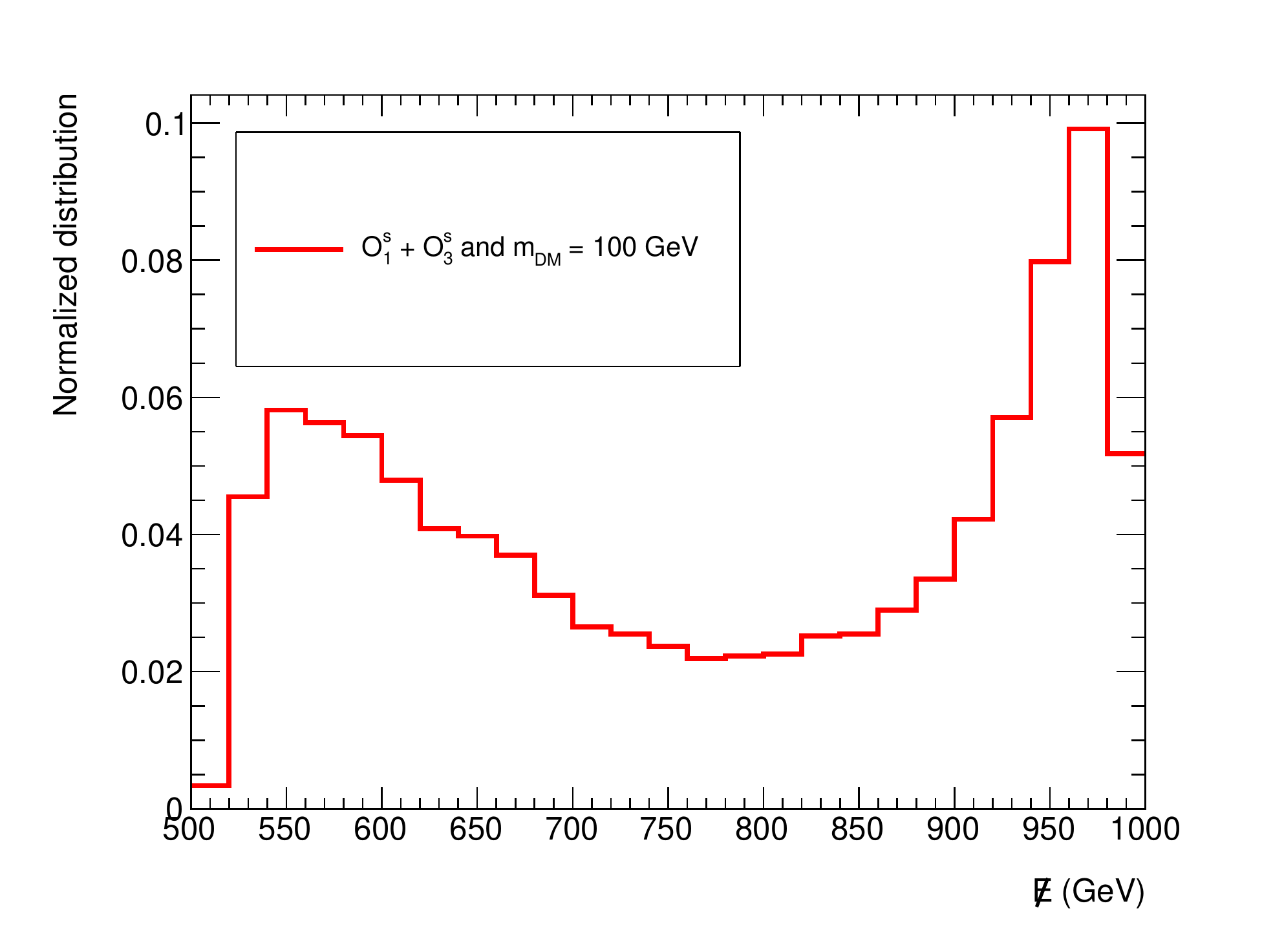}
       	\caption{Double peak behaviour in the mono-$\gamma$ final state, coming from a single DM candidate via two different operators .}
	\label{missenergy_singledm}
\end{figure}

Next we discuss the issue, whether a single WIMP can lead to a double-peak behaviour. There is definitely a possibility there is only one DM candidate 
but it interacts with the SM states with different higher-dimensional operators, which leads to a double peak-like signature, in a way faking a two-component 
behavior. We have checked for our chosen operators this can happen only with mono-photon scenarios and that too with a specific set of operators and DM mass. 
This can happen via two effects submerged together: (i) infra-red divergence which always leads to monotonic increase in $\slashed{E}$, 
whenever the photon is radiated off initial $e^+$ or $e^-$ for operators like $O_1^s, O_2^s, O_1^f, O_2^f$ as in Table I, (ii) for operators $O_3^s$ and $O_3^f$, 
where photon is part of a four-point vertex, there is no divergence and the peak is observed at the lower end of $\slashed{E}$. 
A specific combination of these two types of operators for a specific DM mass (so that the relative cross-sections of the two peaks are of the same order) 
can lead to such fake double-peak. A specific example with $O_1^s$ and $O_3^s$ is shown in Figure~\ref{missenergy_singledm}. 
However, this behaviour is generally absent in mono-$Z$ or mono-Higgs final state, since there is no 
clear divergent and non-divergent contribution to the $\slashed{E}$ distribution in these cases and two non-divergent contributions 
coming from the same DM mass usually do not lead to such fake double peak behaviour.


In this context, one can also ask whether a single WIMP in the presence of a single operator can lead to a double-peak behaviour in the presence of the SM background. 
Since the SM background in this case is monotonically increasing (for mono-$\gamma$ in particular), for a single WIMP interacting via single operator 
which leads to a single peak in the lower $\slashed{E}$ signal distribution may lead to double hump in $\slashed{E}$ distribution, however  
cannot lead to a double-peak in the bin-by-bin $S/\sqrt{B}$ distribution, as analysed here.

\section{Contribution to $\slashed{E}$ from individual DM components}

\begin{figure*}[hptb!]
\begin{minipage}[t]{\columnwidth}
\begin{minipage}[b]{\columnwidth}
\includegraphics[width=\linewidth]{monophoton_o3s_o2f_individual.pdf}\par
\end{minipage}
\end{minipage}\hfill
\begin{minipage}[t]{\columnwidth}
\begin{minipage}[b]{\columnwidth}
\includegraphics[width=\linewidth]{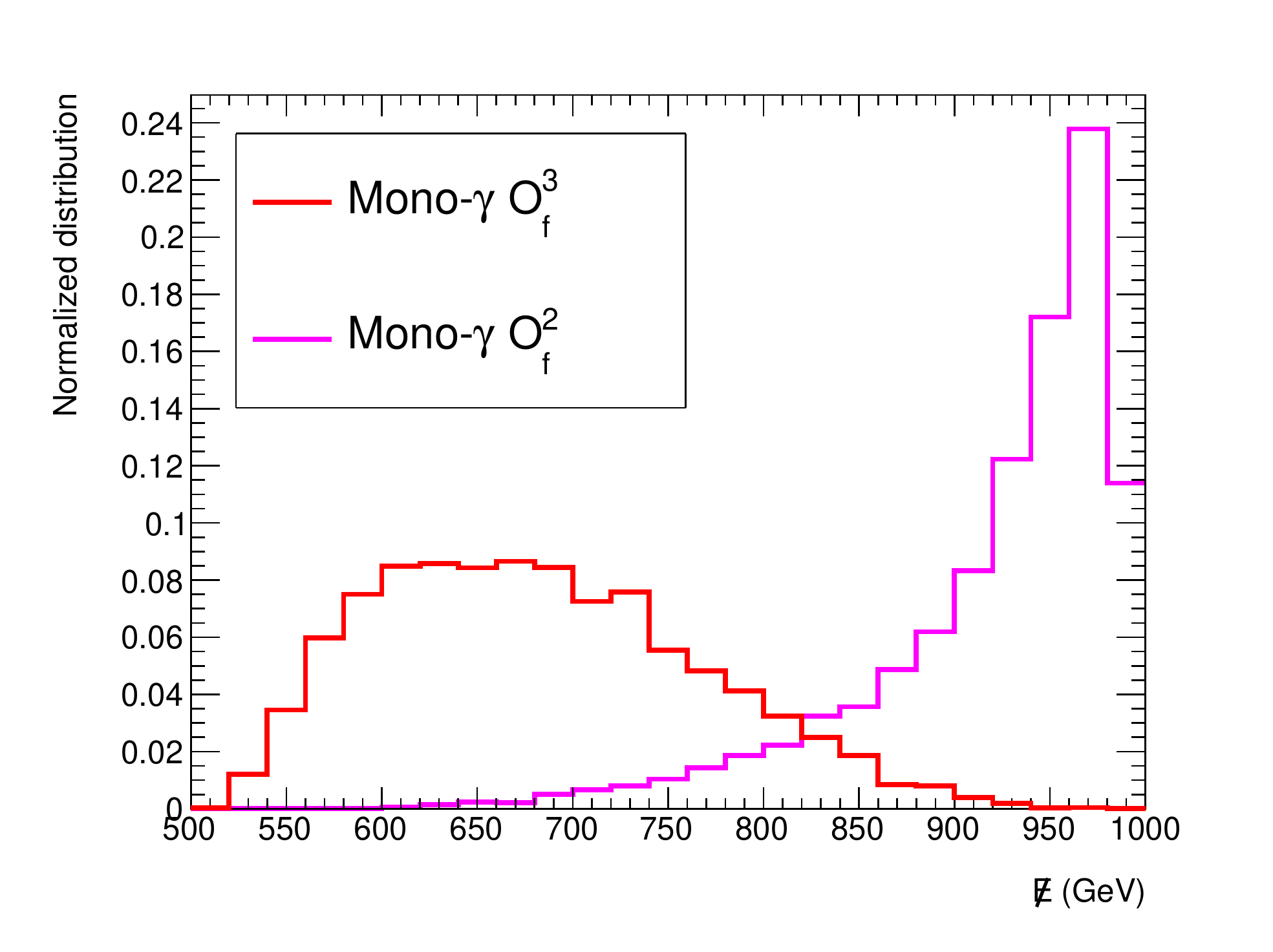}
\end{minipage}
\end{minipage}%
\vspace{0.5cm}
\begin{minipage}[t]{\columnwidth}
\begin{minipage}[b]{\columnwidth}
\includegraphics[width=\linewidth]{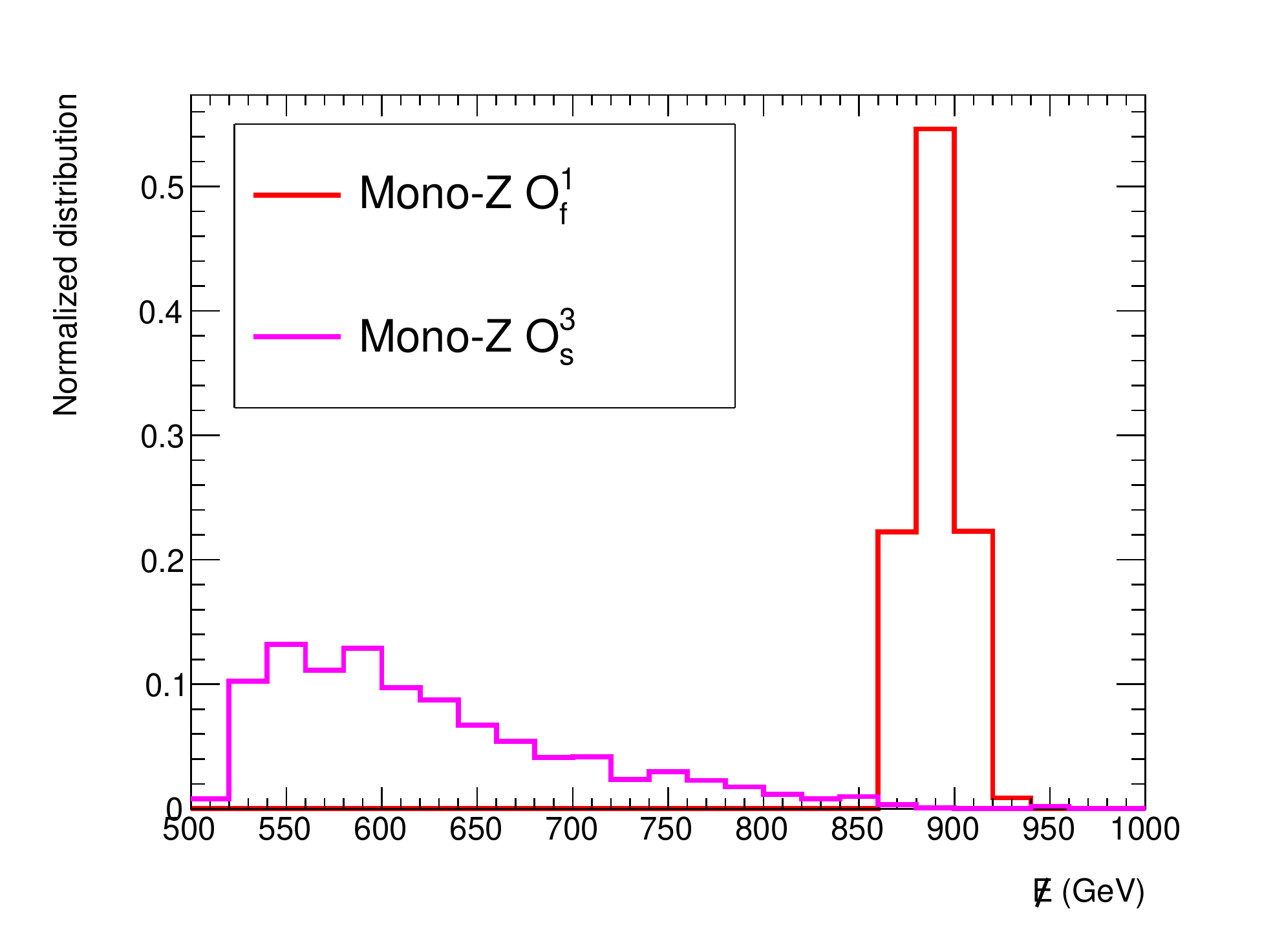}\par
\end{minipage}
\end{minipage}\hfill
\begin{minipage}[t]{\columnwidth}
\begin{minipage}[b]{\columnwidth}
\includegraphics[width=\linewidth]{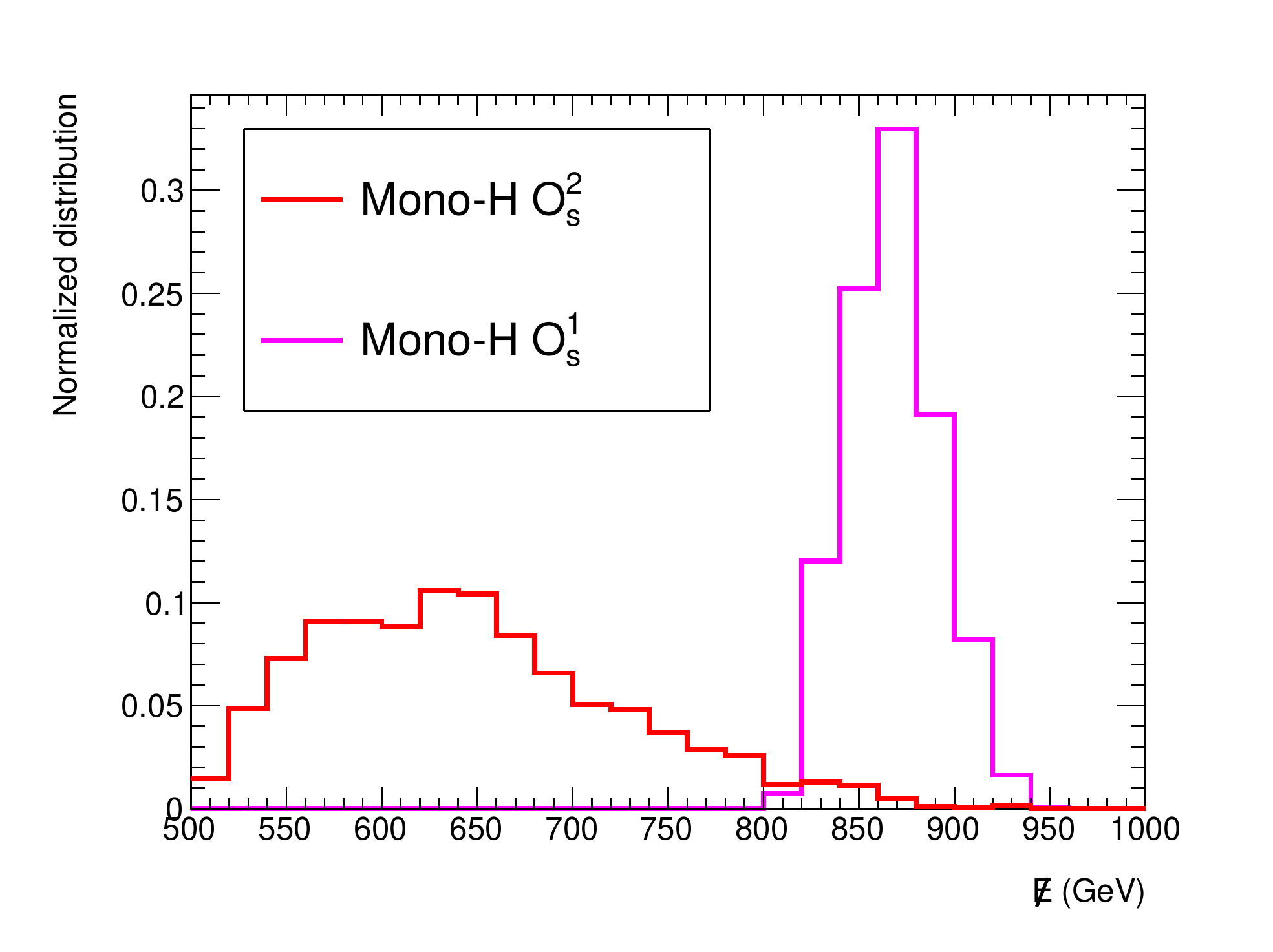}
\end{minipage}
\end{minipage}
\caption{Individual normalized $\slashed{E}$ distributions of various DM candidates represented by the operators (from Table I of the main text) to all the 
	mono-X ($X=\gamma, Z,H$) final states at $e^+e^-$ collisions with $\sqrt{s} = 1$ TeV, $\Lambda=1.2$ TeV and 
	$P_{e^{+}} = +30\%, P_{e^{-}} = +80\%$. The masses of the DMs are chosen according to the benchmark points as in Table III of the draft. See the 
	insets for details.}
\label{missenergy_individual}
\end{figure*}


As elaborated in the text, the double hump distribution in the signal mainly arises for some specific combinations of operators, where 
one gives rise to a peak in the higher $\slashed{E}$, and the other gives rise to a peak in the lower $\slashed{E}$ due to various reasons 
including angular momentum conservation, collinear divergence, rotational invariance and so on. Most often we see that when one operator 
contributes via initial state radiation and the other emits $V$ from the vertex, they combine to give two peak signals. It is often very illuminating to see  
those individual contributions. In Figure~\ref{missenergy_individual}, we show the individual contribution of each operator leading to a double peak 
behaviour for two-DM candidates. The purple and red colours represent two different DM candidates interacting via different operators as described 
in the plot legends. We have shown examples of each of mono-X ($X=\gamma, Z,H$) channels as studied in the draft. 

\section{Relic density contribution from the DM components}

\begin{figure*}[hptb!]
\begin{minipage}[t]{\columnwidth}
\begin{minipage}[b]{\columnwidth}
\includegraphics[width=\linewidth]{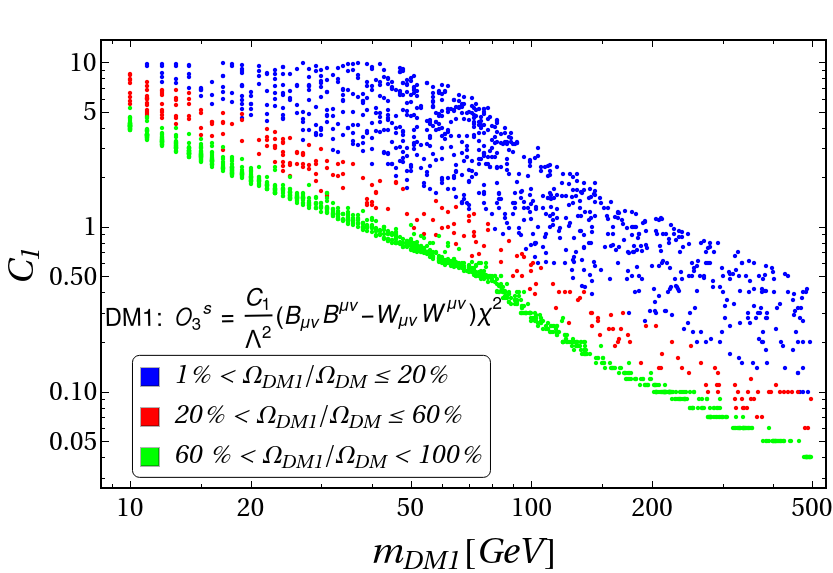}\par
\end{minipage}
\end{minipage}\hfill
\begin{minipage}[t]{\columnwidth}
\begin{minipage}[b]{\columnwidth}
\includegraphics[width=\linewidth]{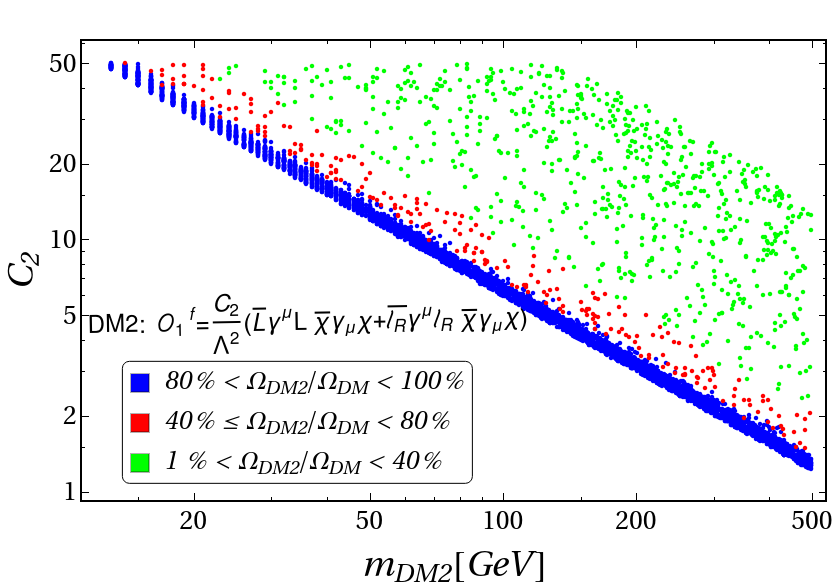}
\end{minipage}
\end{minipage}%
\caption{Relic under abundant parameter space for DM components in $m_{\rm DM1}-C_1$ plane for DM1 which connect to SM via $\mathcal{O}_{3}^s$ (left) and for 
    DM2 in $m_{\rm DM2}-C_2$ plane (right), which connects to SM via operator $\mathcal{O}_{1}^f$. Different colours indicate the percentage of relic density. 
    $\Lambda=1.2$ TeV is used.}
\label{fig:opr}
\end{figure*}

In our work, we have considered DM-SM operators which gives rise to $2_{\rm DM}\to 3_{\rm SM}$ annihilation processes, and 
occasionally $2_{\rm DM}\to 2_{\rm SM}$ ones, giving rise to the relic density for the thermal WIMPS. We calculate the 
freeze-out prospect of the DM components via similar depletion mechanisms, so that each operator produces under-abundance to yield 
$\Omega_{\rm DM}h^2\equiv\Omega_{\rm DM1}h^2+\Omega_{\rm DM2}h^2= 0.120 \pm 0.001$. 
The analysis can be done for each and every combination of DM operators considered in the two component set up, 
however we provide one examples to show how the relic density depends on the ratio of Wilson coefficients to the New Physics scale 
and DM mass. To demonstrate this, we consider two operators in which DM1 and DM2 interact with the SM independently via $\mathcal{O}_3^s$ 
and $\mathcal{O}_1^f $ respectively as 
\begin{eqnarray}
\nonumber 
 {\rm ~DM1:~~~} \mathcal{O}_{3}^s&=&\frac{C_1}{\Lambda^2} \left( B_{\mu\nu}B^{\mu\nu}-W_{\mu\nu}W^{\mu\nu} \right) \chi_{0}^2\, \\
  {\rm ~DM2:~~~}\mathcal{O}_{1}^f&=&\frac{C_2}{\Lambda^2} \left( \overline{L}\gamma^\mu L \overline{\chi_{1/2}}\gamma_\mu \chi_{1/2}  +\overline{\ell_R}\gamma^\nu \ell_R  \overline{\chi_{1/2}}\gamma_\nu \chi_{1/2} \right). 
\end{eqnarray}
Here, $\chi_0$ represents a scalar singlet DM with mass $m_{\rm DM1}$ and $\chi_{1/2}$ represents a fermion singlet DM with mass $m_{\rm DM2}$.  
The relic density of DM1($\chi_0$)  is determined by the parameters $\{ m_{\rm DM1},~C_1,~\Lambda\}$ where as for DM2 ($\chi_{1/2}$) is determined by  $\{ m_{\rm DM2},~C_2,~\Lambda\}$. In our discussion, we used $\Lambda=1.2$ TeV. In Fig.\ref{fig:opr}, we show the correlation among the parameters in $m_{\rm DM1}-C_1$ plane 
(in the left plot) and in $m_{\rm DM2}-C_2$ plane (in the right plot) when both DM satisfy relic under-abundance. Different colours represent the contribution to the relic 
density from DM1 and DM2. We see from these two plots that a possible combination is to choose $m_{\rm DM1}\sim 100 ~\rm GeV, C_1 \sim 1$ to get 
$\Omega_1h^2 \sim 10\%$ while $m_{\rm DM2}\sim 400 ~\rm GeV, C_2 \sim 1.5$ to get $\Omega_2h^2 \sim 90\%$, so that the total relic obeys observed limit. 
This is the combination of masses chosen for producing a double peak signal for mono-Z signal in Fig.~3 of the main text. 

\section{Effect of DM mass in distinguishing the two peak behaviour}

\begin{figure*}[hptb!]
\begin{minipage}[t]{\columnwidth}
\begin{minipage}[b]{\columnwidth}
\includegraphics[width=\linewidth]{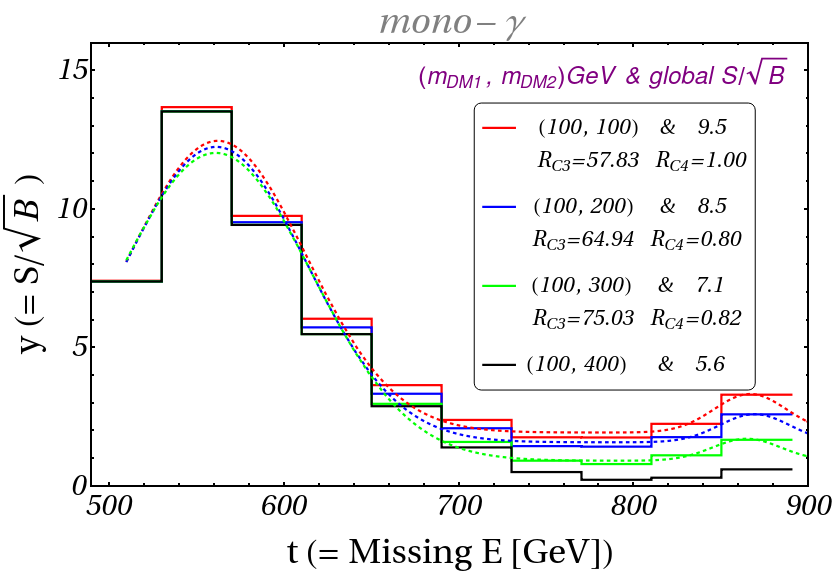}\par
\end{minipage}
\end{minipage}\hfill
\begin{minipage}[t]{\columnwidth}
\begin{minipage}[b]{\columnwidth}
\includegraphics[width=\linewidth]{ss_mp_dm1_100_200_300_400_dm2_100.png}
\end{minipage}
\end{minipage}%
\caption{Local significance distribution in the mono-$\gamma$ final state for two DM components arising from $O_3^s$ (DM1) and $O_2^f$ (DM2) operators 
    considering different combination of DM masses $\{ m_{\rm DM1},~m_{\rm DM2} \}$. $\nu\bar\nu \gamma$ is the dominate SM background here. We choose 
    $\sqrt{s} = 1$ TeV, $\Lambda=1.2$ TeV with $P_{e^{+}} = +30\%, P_{e^{-}} = +80\%$. The corresponding values of the variables $R_{C_3}$ and $R_{C_4}$ are 
    also mentioned in the inset of the figures. Global significance in each combination is also mentioned.}
\label{fig:diffMass}
\end{figure*}

\begin{table}[!hptb]
\begin{center}
\begin{tabular}{| c | c |}
\hline
 $m_{DM1}$(GeV), $m_{DM2}$(GeV) & global $S/\sqrt{B}$  \\
\hline
\hline
100, 100 & 9.5 \\
\hline
100, 200 & 8.5 \\
\hline
100, 300 & 7.1\\
\hline
100, 400 & 5.6 \\
\hline
200, 100 & 6.7 \\
\hline
300, 100 & 5.0\\
\hline
400, 100 & 4.5 \\
\hline
\end{tabular}
\caption{Global significance for mono-$\gamma$ signal for different DM mass combinations from operators $O_3^s$ and $O_2^f$ at $e^+e^-$ collider with $\sqrt s=1$ TeV, 
$\Lambda=1.2$ TeV and $P_{e^{+}} = +30\%, P_{e^{-}} = +80\%$ for integrated luminosity. For background contribution, refer to the main text.}
\label{global}
\end{center}
\end{table}

From the point 5 in the main text, we already know that DM mass plays a crucial role in deciding the lower end of the $\slashed{E}$ spectrum as well the height of the peak 
via production cross-section. We explore the role of DM mass in the distinguishability of two DM components in Figures~\ref{fig:diffMass} (left) and (right). In these figures 
we have considered one particular combination of operators $O_3^s$ (DM1) and $O_2^f$ (DM2). In the left figure mass of DM1 is kept fixed and mass of DM2 is varied. 
In the right figure mass of DM2 is kept fixed and mass of DM1 is varied. We see that with the increase in mass in both cases, the distinguishability of the two peaks worsens. 
This is clear from the variables $R_{C_3}$ and $R_{C_4}$ as mentioned in the figure inset. We also calculate the global significance $S/\sqrt{B}$ for each cases. 
It is seen from Table~\ref{global} that, with the increase in mass the global significance too decreases, due to decrease in the signal production cross-section. 
Therefore it is legitimate to infer that the distinguishability of the two DM components in mono-X signal, not only depends 
on the operator combination, but also on a window of suitable combination of DM masses.

%

\end{document}